\documentclass[longbibliography,twocolumn,preprintnumbers,amsmath,amssymb]{revtex4-1}
\usepackage{graphicx}
\usepackage[utf8,latin1]{inputenc}
\usepackage{dcolumn}
\usepackage{bm}
\usepackage{hyperref}
\usepackage{amsmath}

\usepackage{hyperref}
\hypersetup{colorlinks, linkcolor={red},citecolor={blue},urlcolor={blue}}

\begin{document}

\title[New approaches to the general relativistic Poynting-Robertson effect]{New approaches to the general relativistic Poynting-Robertson effect}

\author{Vittorio De Falco$^{1}$}\email{vittorio.defalco@physics.cz}

\affiliation{$^1$ Research Centre for Computational Physics and Data Processing, Faculty of Philosophy \& Science, Silesian University in Opava, Bezru\v{c}ovo n\'am.~13, CZ-746\,01 Opava, Czech Republic
}

\date{\today}

\begin{abstract}
\emph{Objectives:} A systematic study on the general relativistic Poynting-Robertson effect has been developed so far by introducing different complementary approaches, which can be mainly divided in two kinds: (1) improving the theoretical assessments and model in its simple aspects, and (2) extracting mathematical and physical information from such system with the aim to extend methods or results to other similar physical systems of analogue structure.

\emph{Methods/Analysis:} We use these theoretical approaches: relativity of observer splitting formalism; Lagrangian formalism and Rayleigh potential with a new integration method; Lyapunov theory os stability.

\emph{Findings:} We determined the three-dimensional formulation of the general relativistic Poynting-Robertson effect model. We determine the analytical form of the Rayleigh potential and discuss its implications. We prove that the critical hypersurfaces (regions where there is a balance between gravitational and radiation forces) are stable configurations.

\emph{Novelty /Improvement:} Our new contributions are: to have introduced the three-dimensional description; to have determined the general relativistic Rayleigh potential for the first time in the General Relativity literature; to have provided an alternative, general and more elegant proof of the stability of the critical hypersurfaces.

\end{abstract}

\maketitle

\section{Introduction}
\label{sec:intro}
The last four years have been witness of revolutionary discoveries in astrophysics, which have seen as protagonists these two significant events: (1) the first detection of gravitational waves from the binary black hole (BH) GW151226 \cite{Abott2016a,Abott2016b} and then from the binary neutron star (NS) GW170817 \cite{Abott2017} thanks to the LIGO and VIRGO collaborations; (2) the first imaging of the matter motion around the supermassive BH in the center of M87 Galaxy \cite{EHC20191,EHC20192,EHC20193,EHC20194,EHC20195,EHC20196} thanks to the strong efforts spent on building the Event Horizon Telescope (EHT) and the synergetic cooperation between EHT and Black Hole Cam project. The achievement of such scientific milestones have increasingly motivated all research groups to improve the actual theoretical models, to validate Einstein theory or possible extension of it, when benchmarked with these new amount of powerful observational data. 

Generally, the motion of the matter around massive compact objects, as stellar NSs or BHs, or supermassive BHs, is approximated to be mainly geodetic. However, in view of the actual powerful observational capacities and facilities, it is important to take into account other small perturbing effects. In particular, when we consider the motion of relatively small-sized test particles (e.g., dust grains, gas clouds, meteors, accretion disk matter elements) around electromagnetic radiating sources (like type-I X-ray bursts on NS polar caps, boundary layer around a NS, or a hot corona around a BH) an important effect to be taken into account is the Poynting-Robertson (PR) effect \cite{Poynting1903,Robertson1937}. The forces acting on the test particle are: the gravitational field, directed toward the compact object and opposite to the radiation pressure, pointing outward, and also the PR effect. This phenomenon is triggered each time the radiation field invests the test particle, raising up its temperature, which for the Stefan-Boltzmann law starts re-emitting radiation. In this model, the test particle is considered as an ideal black body in thermal equilibrium, meaning that all the absorbed energy is isotropically re-emitted. 

This process of absorption and remission of radiation generates a recoil force opposite to the test body orbital motion. This can be interpreted as an aberration effect in the test particle's frame or also as an anisotropic re-emission in the star reference frame. It is important to note that radiation pressure and PR effect can be split in the classical frame, while in GR frame they constitute one single function, which must satisfy the relativistic covariance principle, in order not to run into paradoxes. However, the PR effect can be seen as the action of an electromagnetic field on a moving body. Such mechanism removes thus very efficiently angular momentum and energy from the test particle, forcing it to spiral inward or outward depending on the radiation field intensity. 

Such effect has been initially introduced in classical physics by Poynting in 1903 \cite{Poynting1903}, and then extended in special relativity by Robertson in 1937 \cite{Robertson1937}, with several applications to the Solar system \cite{Burns1979}. From 2009, it has been extended in GR by Bini and collaborators within the two-dimensional (2D) equatorial plane of the Kerr spacetime \cite{Bini2009,Bini2011}. Recently, it has been extended also in the three-dimensional (3D) space \cite{Defalco20183D,Bakala2019,Wielgus2019}. 

Our aim is to investigate such phenomenon under different perspectives, in order to extract more peculiar information, and for developing new valuable mathematical tools, which can be broadly applied to other dissipative systems in GR. in this paper, we would like to revise our new approaches and the consequent implications. The article is structured as follows: in Sec. \ref{sec:3D}, we concentrate on the PR effect model, showing how to pass from the 2D description to the 3D case, discussing the further implications and new advantages; in Sec. \ref{sec:RF}, we treat the PR effect as a dissipative system under a Lagrangian formalism, determining analytically the Rayleigh dissipation function through a new procedure; in Sec. \ref{sec:stbch}, we study the PR effect as a dynamical system, proving that the critical hypersurfaces (regions where gravitational and radiation forcese balance) are stable configurations within Lyapunov theory. Finally, in Sec. \ref{sec:end} the conclusions are drawn.

\section{From the 2D to 3D general relativistic PR effect model}
\label{sec:3D}
\begin{figure*}[t!]
	\centering
	\hbox{
	\includegraphics[scale=0.45]{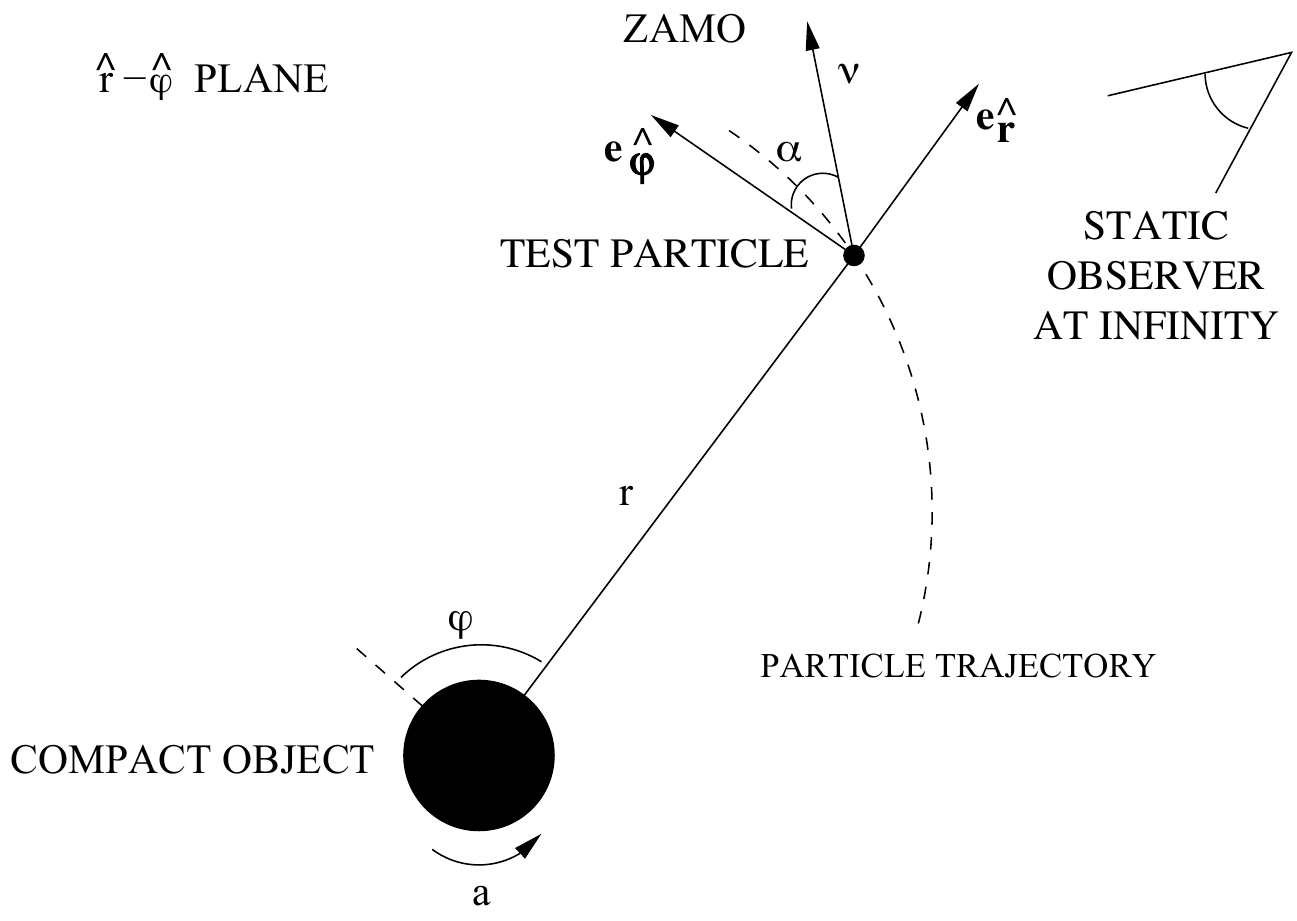}
	\hspace{0.2cm}
        \includegraphics[scale=0.32]{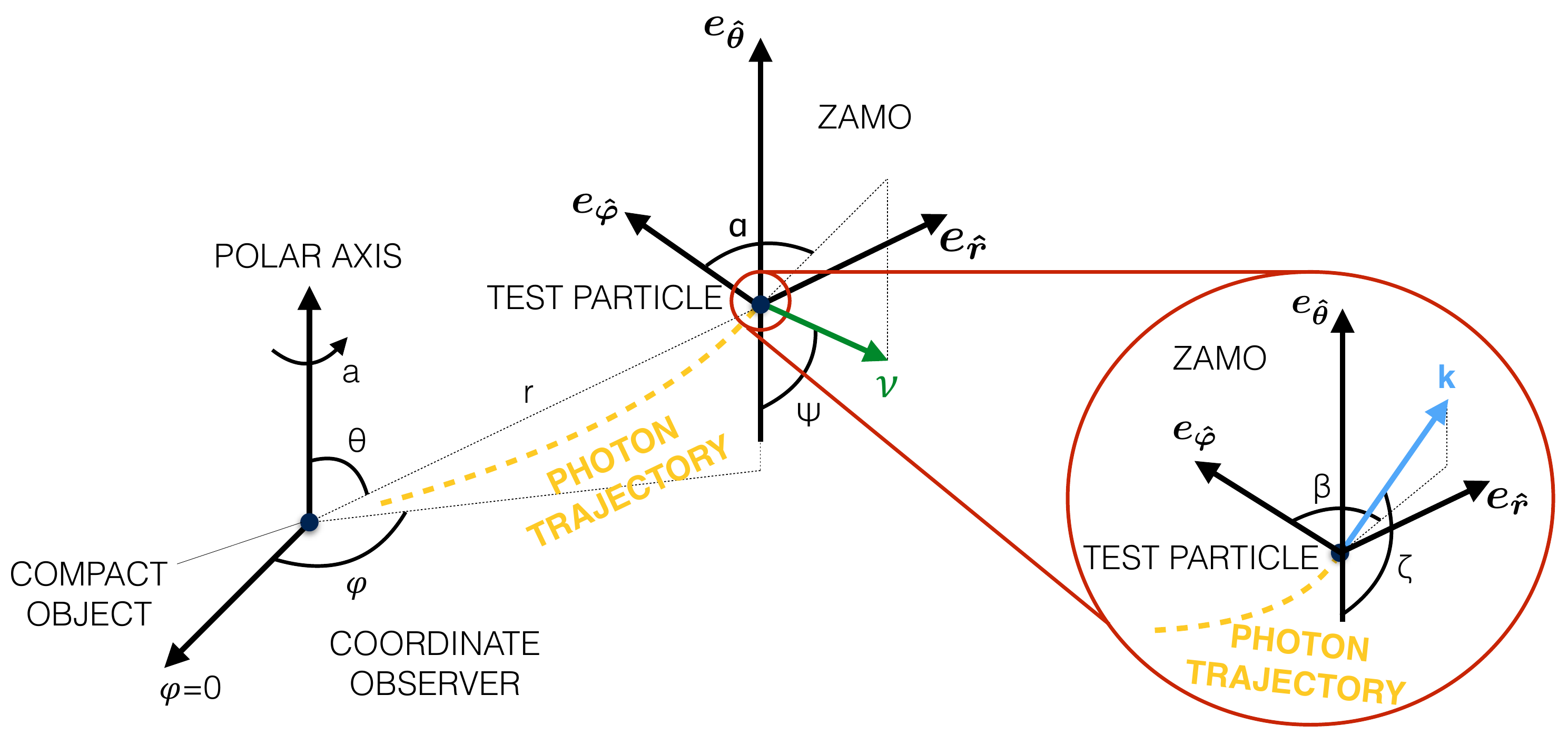}}
	\caption{Geometries of the 2D (left panel) and 3D general relativistic PR effect models (right panel).}
	\label{fig:Plot1}
\end{figure*}

\subsection{Geometry and strategy}
We consider a rotating compact object, whose outside spacetime is described by the Kerr metric. We use the signature $(-,+,+,+)$ for the metric, and geometrical units for gravitational constant $G$, and speed of light $c$ ($c = G = 1$). The metric line element, $ds^2=g_{\alpha\beta}dx^\alpha dx^\beta$, expressed in Boyer-Lindquist coordinates, parameterized by mass $M$ and spin $a$, reads as \cite{Misner1973}
\begin{equation}\label{kerr_metric}
\begin{aligned}
 \mathrm{d}s^2 &= \left(\frac{2Mr}{\Sigma}-1\right)\mathrm{d}t^2 
  - \frac{4Mra\sin^2\theta}{\Sigma}\mathrm{d}t \mathrm{d}\varphi+ \frac{\Sigma}{\Delta}\mathrm{d}r^2 \\
  &+ \Sigma\mathrm{d}\theta^2 + \rho\sin^2\theta\mathrm{d}\varphi^2, 
\end{aligned}
\end{equation}
where $\Sigma \equiv r^{2} + a^{2}\cos^{2}\theta$, $\Delta \equiv r^{2} - 2Mr + a^{2}$, and $\rho  \equiv r^2+a^2+2Ma^2r\sin^2\theta/\Sigma$. 

The strategy for determining the equations of motion of a test particle under the general relativistic PR effect is obtained by following the initial approach followed by Poynting and Robertson. We write first the equations in the test particle rest frame through the zero angular momentum observers (ZAMOs), considered at the position occupied by the test particle at each instant of time. Then, we transform them in the static observer frame located at infinity. The successful technique exploited to achieve such objective is reached through the \emph{relativity of observer splitting formalism}. This represents a powerful method in GR to distinguish the gravitational effects from the fictitious forces arising from the relative motion of two non-inertial observers \cite{Jantzen1992,Bini1997a,Bini1997b,Defalco2019}. Such formalism allows us to derive the test particle equations of motion in the reference frame of the static observer located at infinity as a set of coupled first order ordinary and highly-linear differential equations \cite{Bini2009,Bini2011,Defalco20183D,Bakala2019}.

The orthonormal frame adapted to the ZAMOs is \cite{Bini2009,Bini2011,Defalco20183D,Bakala2019}
\begin{equation} \label{eq:zamoframes}
\begin{aligned}
&\boldsymbol{e_{\hat t}}\equiv\boldsymbol{n}= \frac{(\boldsymbol{\partial_t}-N^{\varphi}\boldsymbol{\partial_\varphi})}{N},\quad \boldsymbol{e_{\hat r}}=\frac{\boldsymbol{\partial_r}}{\sqrt{g_{rr}}},\\
&\boldsymbol{e_{\hat \theta}}=\frac{\boldsymbol{\partial_\theta}}{\sqrt{g_{\theta \theta }}},\quad \boldsymbol{e_{\hat \varphi}}=\frac{\boldsymbol{\partial_\varphi}}{\sqrt{g_{\varphi \varphi }}}.
\end{aligned}
\end{equation}
where $N=(-g^{tt})^{-1/2}$ and $N^{\varphi}=g_{t\varphi}/g_{\varphi\varphi}$. The nonzero ZAMO kinematical quantities in the decomposition of the ZAMO congruence are acceleration $\boldsymbol{a}(n)=\nabla_{\boldsymbol{n}} \boldsymbol{n}$, expansion tensor along the $\hat{\varphi}$-direction $\boldsymbol{\theta_{\hat\varphi}}(n)$, and the relative Lie curvature vector $\boldsymbol{k_{(\rm Lie)}}(n)$ (see Table 1 in \cite{Defalco20183D}, for their explicit expressions). We denote scalar and tensors measured in the ZAMO frame respectively followed by $(n)$ and by a superposed hat. 

\subsection{Radiation field}
The radiation field is modeled as a coherent flux of photons traveling along null geodesics on the Kerr metric. We consider that at each instant of time a single photon, from an emitting surface around the central compact object, reaches the test particle in its position. The related stress-energy tensor $T^{\mu\nu}$ is given by \cite{Defalco20183D,Bakala2019}
\begin{equation}\label{STE}
T^{\mu\nu}=\mathcal{I}^2 k^\mu k^\nu\,,\qquad k^\mu k_\mu=0,\qquad k^\mu \nabla_\mu k^\nu=0,
\end{equation}
where $\mathcal{I}$ is a parameter linked to the radiation field intensity and $\boldsymbol{k}$ is the photon four-momentum field, where the last two equations express the condition of null geodesic.  In Kerr spacetime, we have that the energy $E=-k_t$, the angular momentum with respect to the polar axis $L_z=k_\varphi$, the Carter constant $\mathcal{Q}$, and the module of the photons $k_\mu k^\mu=0$ are conserved along its trajectory.

Splitting $\boldsymbol{k}$ with respect to the ZAMO frame (see Fig. \ref{fig:Plot1}), we obtain \cite{Bini2009,Bini2011,Defalco20183D,Bakala2019}
\begin{eqnarray}
&&\boldsymbol{k}=E(n)[\boldsymbol{n}+\boldsymbol{\hat{\nu}}(k,n)], \label{photon1}\\
&&\boldsymbol{\hat{\nu}}(k,n)=\sin\zeta\sin\beta\ \boldsymbol{e_{\hat r}}+\cos\zeta\ \boldsymbol{e_{\hat\theta}}+\sin\zeta \cos\beta\ \boldsymbol{e_{\hat\varphi}}, \label{photon2}
\end{eqnarray}
where $E(n)$ is the photon energy measured in the ZAMO frame, $\boldsymbol{\hat{\nu}}(k,n)$ is the photon spatial unit relative velocity with respect to the ZAMOs, $\beta$ and $\zeta$ are the two angles measured in the ZAMO frame in the azimuthal and polar direction, respectively. The radiation field in the 3D model is governed by the two impact parameters $(b,q)$, associated respectively with the two emission angles $(\beta,\zeta)$. The radiation field photons are emitted from a spherical rigid surface having a radius $R_\star$ centered at the origin of the Boyer-Lindquist coordinates, and rotating rigidly with angular velocity $\Omega_{\mathrm{\star}}$. In the 2D model, the motion occurs only in the equatorial plane $\theta=\pi/2$, where we have only one impact parameter $b$ related to the only emission angle $\beta$ in the ZAMO frame. We say that for $b=0$ the radiation field is radial, otherwise it is a general radiation field (see Fig. \ref{fig:Plot1}). 

The photon impact parameters are \cite{Bini2011,Bakala2019}
\begin{eqnarray} 
&&b\equiv \frac{L}{E}=-\left[\frac{\mathrm{g_{t\varphi}}+\mathrm{g_{\varphi\varphi}}\Omega_{\star} }{\mathrm{g_{tt}}+\mathrm{g_{t\varphi}} \Omega_{\star}}\right]_{r=R_\star},\label{kerr_impact_parameter}\\
&&q\equiv \frac{\mathcal{Q}}{E^2}=\left[b^{2}\cot^{2} \theta-a^{2} \cos^{2}\theta \right]_{r=R_\star}. \label{q_r}
\end{eqnarray}
Both relations are evaluated at the emitting surface radius $R_\star$. For the azimuthal photon impact parameter $b$, we have that $b=b(R_\star,\Omega_\star,\theta,a)$, where $\theta$ is the polar angle occupied by the test particle during its motion. Instead, for the latitudinal photon impact parameter $q$, it fixes the Carter constant $\mathcal{Q}$ for a given
photon trajectory, and depends only on the test particle's polar angle $\theta$ and the value assumed by $b$.

The related photon angles in the ZAMO frame are \cite{Bakala2019}
\begin{equation} \label{ANG1}
\cos\beta=\frac{b N}{\sqrt{g_{\varphi\varphi}}(1+b N^\varphi)}, \qquad \zeta=\pi/2. 
\end{equation}
By imposing that the photon must hit the test particle in the equatorial plane of the ZAMO frame, even at infinity, we substantially reduce the complexity of the problem, because everything is expressed in terms of one single parameter, which is astrophysically realistic.

In both 2D and 3D models, from the conservation of the stress-energy tensor conditions (Bianchi identities), namely $\nabla_\mu T^{\mu\nu}=0$, we are able to determine the parameter $\mathcal{I}$, which has the following expression \cite{Bakala2019}
\begin{equation}\label{INT_PAR}
\mathcal{I}^2=\frac{\mathcal{I}_0^2}{\sqrt{\left( r^{2} + a^{2}-ab \right)^{2}- \Delta \left[ q + \left( b - a \right) ^{2} \right]}},
\end{equation}
where $\mathcal{I}_0$ is $\mathcal{I}$ evaluated at the emitting surface.

\subsection{Equations of motion}
A test particle moves with a timelike four-velocity $\boldsymbol{U}$ and a spatial three-velocity with respect to the ZAMO frames, $\boldsymbol{\nu}(U,n)$, which both read as (see Fig. \ref{fig:Plot1}) \cite{Bakala2019}
\begin{eqnarray} 
&&\boldsymbol{U}=\gamma(U,n)[\boldsymbol{n}+\boldsymbol{\nu}(U,n)], \label{testp}\\
&&\boldsymbol{\nu}=\nu(\sin\psi\sin\alpha\boldsymbol{e_{\hat r}}+\cos\psi\boldsymbol{e_{\hat\theta}}+\sin\psi \cos\alpha \boldsymbol{e_{\hat\varphi}}),
\end{eqnarray}
where $\gamma(U,n)\equiv\gamma=1/\sqrt{1-||\boldsymbol{\nu}(U,n)||^2}$ is the Lorentz factor, $\nu=||\boldsymbol{\nu}(U,n)||$, $\gamma(U,n) =\gamma$. We have that $\nu$ represents the magnitude of the test particle spatial velocity $\boldsymbol{\nu}(U,n)$, $\alpha$ is the azimuthal angle of the vector $\boldsymbol{\nu}(U,n)$ measured clockwise from the positive $\hat\varphi$ direction in the $\hat{r}-\hat{\varphi}$ tangent plane in the ZAMO frame, and $\psi$ is the polar angle of the vector $\boldsymbol{\nu}(U,n)$ measured from the axis orthogonal to the $\hat{r}-\hat{\varphi}$ tangent plane in the ZAMO frame. 

We assume that the radiation-test particle interaction occurs through Thomson scattering, characterized by a constant momentum-transfer cross section $\sigma$, independent from direction and frequency of the radiation field. We can split the photon four momentum (\ref{photon1}) in terms of the velocity $\boldsymbol{U}$ as \cite{Bakala2019}
\begin{equation}
\boldsymbol{k}=E(U)[\boldsymbol{U}+\boldsymbol{\hat{\mathcal{V}}}(k,U)],
\end{equation}
where $E(U)$ is the photon energy measured by the test particle. The radiation force can be written as \cite{Bakala2019}
\begin{equation} \label{radforce}
\begin{aligned}
{\mathcal F}_{\rm (rad)}(U)^{\hat \alpha}&\equiv-\tilde{\sigma\mathcal{I}}^2(T^{\hat \alpha}{}_{\hat \beta} U^{\hat \beta}+U^{\hat \alpha} T^{\hat \mu}{}_{\hat \beta} U_{\hat \mu} U^{\hat \beta})\\
&=\tilde{\sigma} \, [\mathcal{I} E(U)]^2\, \hat{\mathcal V}(k,U)^{\hat \alpha},
\end{aligned}
\end{equation}
where $m$ is the test particle mass and the term $\tilde{\sigma}[\mathcal{I} E(U)]^2$ reads as \cite{Bakala2019} 
\begin{equation} \label{eq: sigma_tilde}
\tilde{\sigma}[\mathcal{I} E(U)]^2=\frac{ A\,\gamma^2(1+b N^\varphi)^2[1-\nu\sin\psi\cos(\alpha-\beta)]^2}{N^2\sqrt{R(r)}},
\end{equation}
where
\begin{equation}
R(r)=\left( r^{2} + a^{2}-ab \right)^{2}- \Delta \left[ q + \left( b - a \right) ^{2} \right],
\end{equation}
and with $A=\tilde{\sigma}[\mathcal{I}_0 E]^2$ being the luminosity parameter, which can be equivalently written as $A/M=L/L_{\rm EDD}\in[0,1]$ with $L$ the emitted luminosity at infinity and $L_{\rm EDD}$ the Eddington luminosity. The terms $\hat{\mathcal V}(k,U)^{\hat \alpha}$ are the radiation field components, whose expressions are \cite{Bakala2019}
\begin{eqnarray}\label{rad}
&&\hat{\mathcal{V}}^{\hat r}=\frac{\sin\beta}{\gamma [1-\nu\sin\psi\cos(\alpha-\beta)]}-\gamma\nu\sin\psi\sin\alpha, \\ &&\hat{\mathcal{V}}^{\hat \theta}=-\gamma\nu\cos\psi,\\
&&\hat{\mathcal{V}}^{\hat\varphi}=\frac{\cos\beta}{\gamma [1-\nu\sin\psi\cos(\alpha-\beta)]}-\gamma\nu\sin\psi\cos\alpha,\\
&&\hat{\mathcal{V}}^{\hat t}=\gamma\nu\left[\frac{\sin\psi\cos(\alpha-\beta)-\nu}{1-\nu\sin\psi\cos(\alpha-\beta)}\right].
\end{eqnarray}

Gathering all information together, it is possible to derive the resulting equations of motion for a test particle moving in a 3D space, which are \cite{Bini2009,Bini2011,Defalco20183D,Bakala2019}
\begin{eqnarray}
&&\frac{d\nu}{d\tau}= -\frac{1}{\gamma}\left\{ \sin\alpha \sin\psi\left[a(n)^{\hat r}\right.+2\nu\cos \alpha\sin\psi\, \theta(n)^{\hat r}{}_{\hat \varphi} \right]\label{EoM1}\\
&&\left.+\cos\psi\left[a(n)^{\hat \theta}+2\nu\cos\alpha\sin\psi\, \theta(n)^{\hat \theta}{}_{\hat \varphi}\right]\right\}+\frac{\tilde{\sigma}[\Phi E(U)]^2}{\gamma^3\nu}\hat{\mathcal{V}}^{\hat t},\nonumber\\
&&\frac{d\psi}{d\tau}= \frac{\gamma}{\nu} \left\{\sin\psi\left[a(n)^{\hat \theta}+k_{\rm (Lie)}(n)^{\hat \theta}\,\nu^2 \cos^2\alpha\right.\right.\label{EoM2}\\
&&\left.\left.+2\nu\cos \alpha\sin\psi\ \theta(n)^{\hat \theta}{}_{\hat \varphi}\right]-\sin \alpha\cos\psi \left[a(n)^{\hat r}+k_{\rm (Lie)}(n)^{\hat r}\,\nu^2\right.\right.\nonumber\\
&&\left.\left.+2\nu\cos \alpha\sin\psi\, \theta(n)^{\hat r}{}_{\hat \varphi}\right]\right\}+\frac{\tilde{\sigma}[\Phi E(U)]^2}{\gamma\nu^2\sin\psi}\left[\hat{\mathcal{V}}^{\hat t}\cos\psi-\hat{\mathcal{V}}^{\hat \theta}\nu\right],\nonumber\\
&&\frac{d\alpha}{d\tau}=-\frac{\gamma\cos\alpha}{\nu\sin\psi}\left[a(n)^{\hat r}+2\theta(n)^{\hat r}{}_{\hat \varphi}\ \nu\cos\alpha\sin\psi\right.\label{EoM3}\\
&&\left.+k_{\rm (Lie)}(n)^{\hat r}\,\nu^2+k_{\rm (Lie)}(n)^{\hat \theta}\,\nu^2\cos^2\psi \sin\alpha\right]\nonumber\\
&&+\frac{\tilde{\sigma}[\Phi E(U)]^2\cos\alpha}{\gamma\nu\sin\psi}\left[\hat{\mathcal{V}}^{\hat r}-\hat{\mathcal{V}}^{\hat \varphi}\tan\alpha\right],\nonumber\\
&&U^{\hat r}\equiv\frac{dr}{d\tau}=\frac{\gamma\nu\sin\alpha\sin\psi}{\sqrt{g_{rr}}}, \label{EoM4}\\
&&U^{\hat \theta}\equiv\frac{d\theta}{d\tau}=\frac{\gamma\nu\cos\psi}{\sqrt{g_{\theta\theta}}} \label{EoM5},\\
&&U^{\hat \varphi}\equiv\frac{d\varphi}{d\tau}=\frac{\gamma\nu\cos\alpha\sin\psi}{\sqrt{g_{\varphi\varphi}}}-\frac{\gamma N^\varphi}{N},\label{EoM6}\\
&&U^{\hat t}\equiv \frac{dt}{d\tau}=\frac{\gamma}{N},\label{time}
\end{eqnarray}
where $\tau$ is the affine parameter (proper time) along the test particle trajectory. Naturally, these equations reduce to the 2D case, when $\psi=\theta=\pi/2$.

\subsection{Critical hypersurfaces and test particle's trajectories}
\label{sec:critc_rad}
\begin{figure*}[t!]
	\centering
	\includegraphics[scale=0.88]{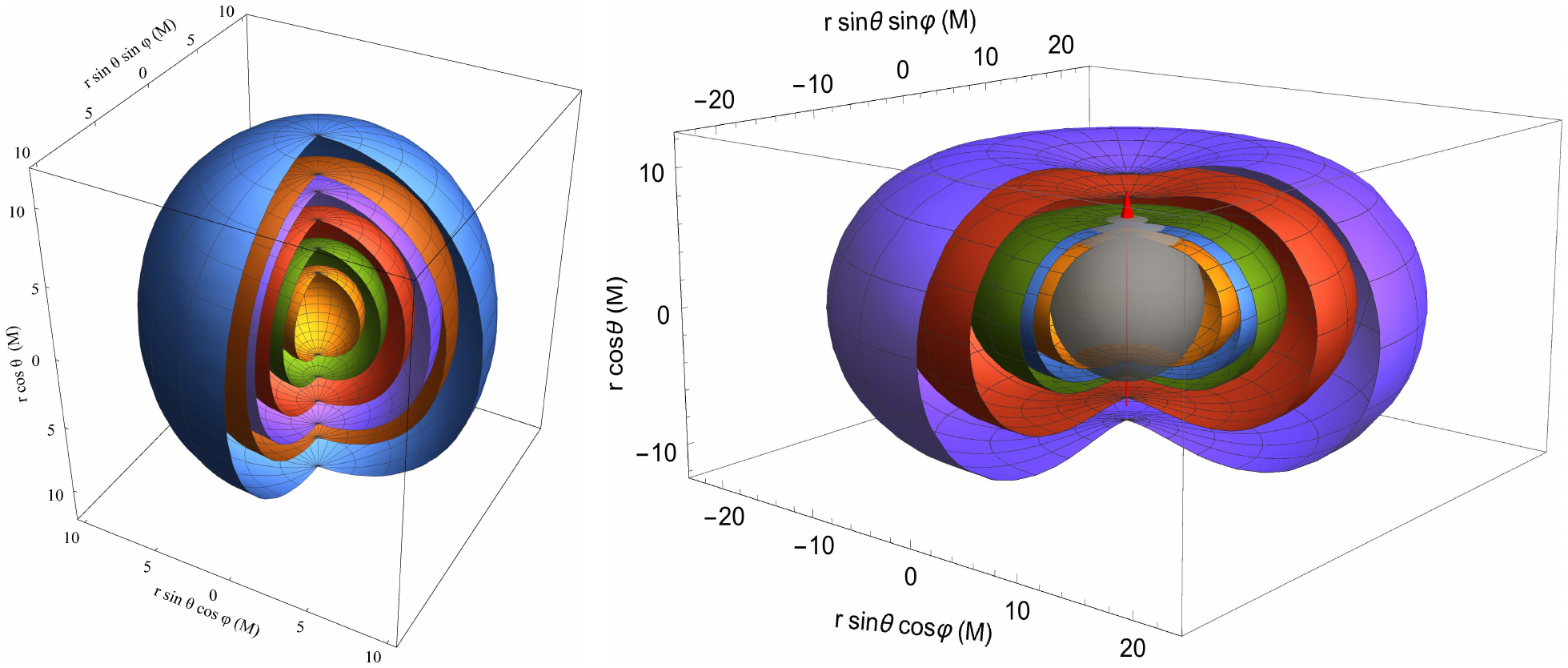}
	\caption{Left panel: Critical hypersurfaces for $\Omega_\star=0$ and the luminosity parameters $A=0.5, \,0.7, \,0.8, \,0.85, \,0.87, \,0.9$ at a constant spin $a=0.9995$. The respective critical radii in the equatorial plane are $r^{\rm eq}_{\rm(crit)} \sim 2.71M,  4.01M, 5.52M, 7.04M, 7.99M, 10.16M$, while at poles they are $r^{\rm pole}_{\rm(crit)} \sim 2.97M,  4.65M,  6.56M,  8.38M,  9.48M, 11.9M$.
	Right panel: Critical hypersurfaces for a NS (grey sphere) with $\Omega_\star=0.031$, $R_\star=6M$, and luminosity parameters $A=0.75,\, 0.78, \,0.8, \,0.85, \,0.88$ at a constant spin $a=0.41$. The respective critical radii in the equatorial plane are $r^{\rm eq}_{\rm(crit)} \sim 8.88M,\  10.61M,\ 12.05M,\ 17.26M,\ 22.43M,\ $, while at poles they are $r^{\rm pole}_{\rm(crit)} \sim 4.73M,\  5.28M,\  5.74M,\  7.43M,\  9.11M$. The red arrow is the polar axis.}
	\label{fig:Fig1}
\end{figure*}

The general relativistic PR effect, defined by Eqs. (\ref{EoM1})--(\ref{EoM6}), exhibits, both in the 2D and 3D models, a critical hypersurface around the compact object. This is a region, where there exists a balance among gravitational and radiation forces, see Fig. \ref{fig:Fig1} for some examples. 
Such structures are obtained by the requirement that the test particle moves on them ($\alpha=0,\pi$) with constant velocity ($\nu=\mbox{const}$) with respect to the ZAMO frame, and the polar axis is orthogonal to the critical hypersurface ($\psi=\pm\pi/2$), which in turn implies that $d\nu/d\tau=d\alpha/d\tau=0$ \cite{Bini2009,Bini2011,Defalco20183D,Bakala2019}
\begin{eqnarray}
&&\nu=\cos\beta, \label{eq:crit_hyper1} \\
&&a(n)^{\hat r}+2\theta(n)^{\hat r}{}_{\hat\varphi}\nu+k_{\rm (Lie)}(n)^{\rm \hat r}\\
&&=\frac{A(1+bN^\varphi)^2\sin^3\beta}{N^2\gamma\sqrt{R(r_{\rm (crit)})}}\,\nu^2\notag\label{eq:crit_hyper2}
\end{eqnarray}
where the first condition means that the test particle moves on the critical hypersurface with constant velocity equal to the azimuthal photon velocity, see Eq. (\ref{photon2}); whereas the second condition determine through an implicit equation the shape of the critical hypersurface in terms of the critical radius $r_{\rm (crit)}$ as a function of the polar angle, once metric background (i.e., $a$) and radiation field proprieties (i.e., $A,R_\star,\Omega_\star$) are assigned. 

When the test particle reaches the critical hypersurface, we can have to possible behaviors: 
\begin{itemize}
\item \emph{latitudinal drift}, due to the interplay of gravitational and radiation actions in the polar direction, which brings definitively the test particle on the equatorial plane \cite{Defalco20183D,Bakala2019}. This corresponds to the condition $d\psi/d\tau\neq0$, because the $\psi$ angle change during the test particle motion on the critical hypersurface; 

\item \emph{suspended orbits}, the test particle does not drift down to the equatorial plane, but moves on a pure circular orbit at $\theta=\bar{\theta}\neq\pi/2$. This condition is mathematically achieved by imposing that $d\psi/d\tau=0$, which for $b\neq0$ reads as \cite{Bakala2019}
\begin{equation}\label{eq:susporbit}
\begin{aligned}
&a(n)^{\hat \theta}+k_{\rm (Lie)}(n)^{\hat \theta}\,\nu^2+2\nu\sin\psi\ \theta(n)^{\hat \theta}{}_{\hat \varphi}\\
&+\frac{A (1+bN^\varphi)^2(1-\cos^2\beta\sin\psi)\cos\beta}{\gamma N^2\sqrt{R(r_{\rm (crit)})}\tan\psi}=0,
\end{aligned}
\end{equation} 
which is an implicit equation in terms of $\psi$. Instead for $b=0$ we obtain $\psi=\pm\pi/2$ \cite{Defalco20183D}. It is important to note that in the Schwarzschild case ($a=0$) for $b=0$ the test particle stops on a point on the critical hypersurface, without moving on a purely circular orbit \cite{Defalco20183D}.

\end{itemize}

\begin{figure*}[t!]
\hspace{6cm}
	\vbox{	\centering\hbox{
	\includegraphics[scale=0.25]{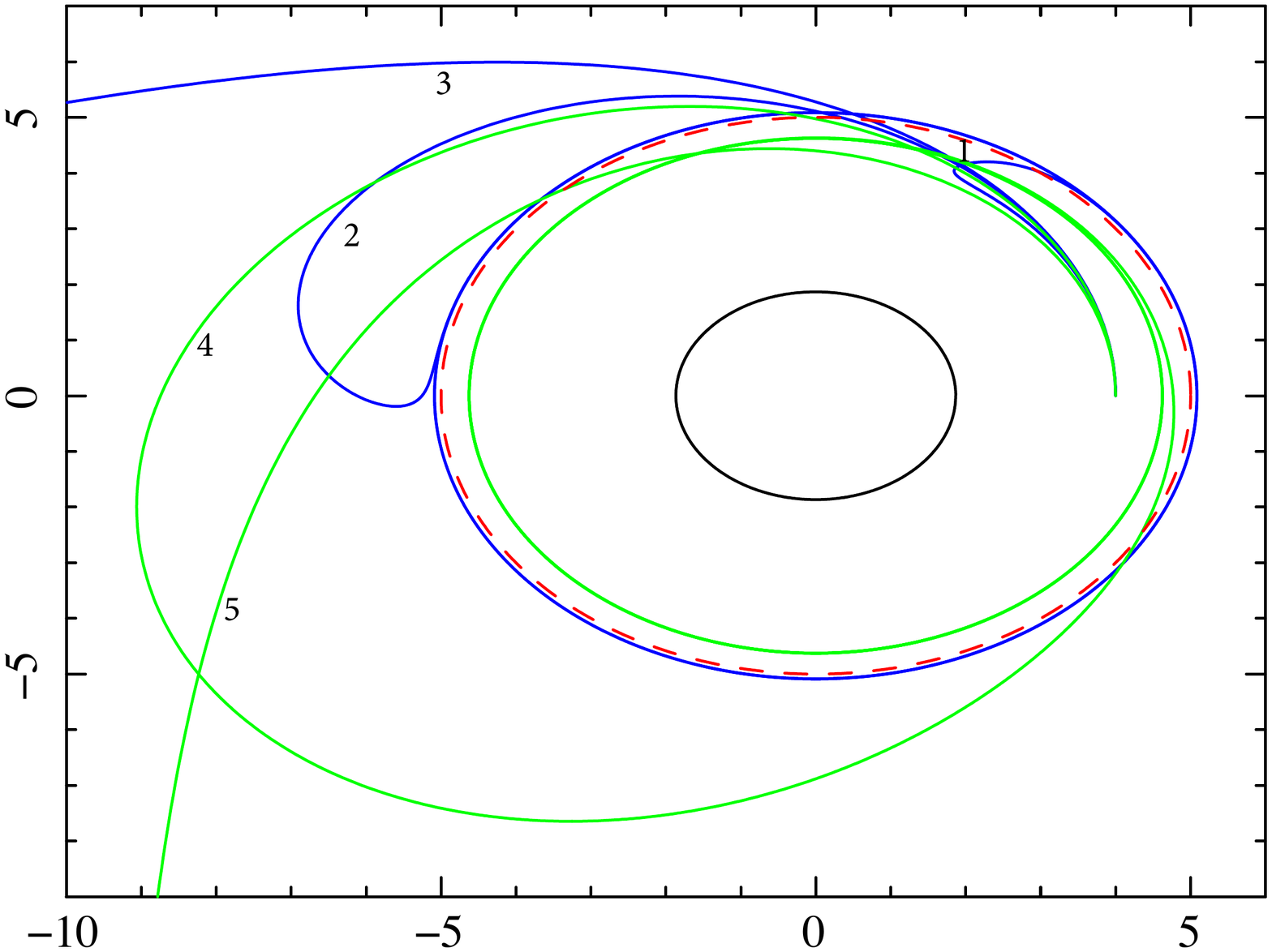}
	\hspace{-0.8cm}
	\includegraphics[scale=0.25]{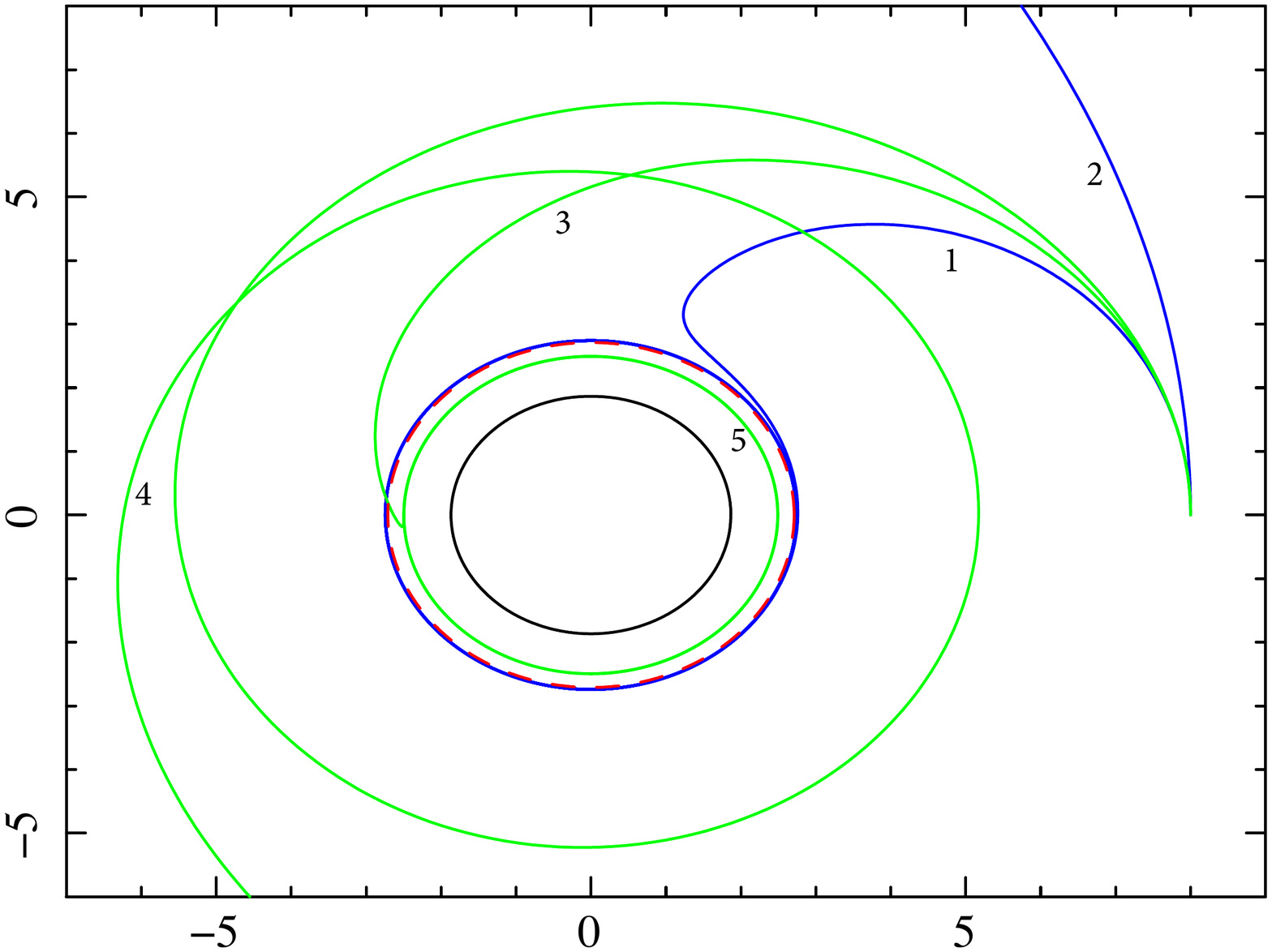}}
	\vspace{0.2cm}
	\hbox{
	\includegraphics[scale=0.25]{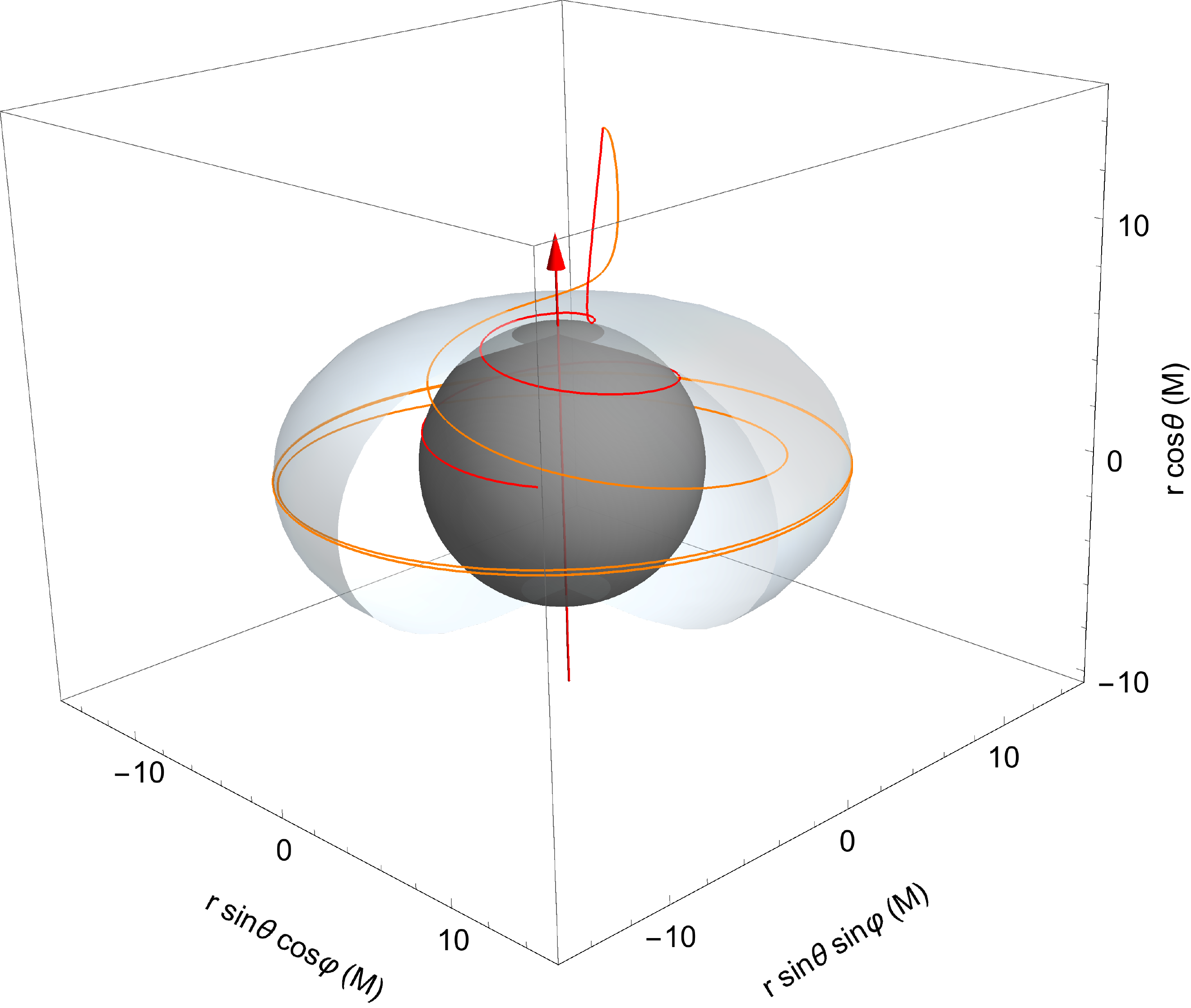}
	\hspace{0.5cm}
	\includegraphics[scale=0.2]{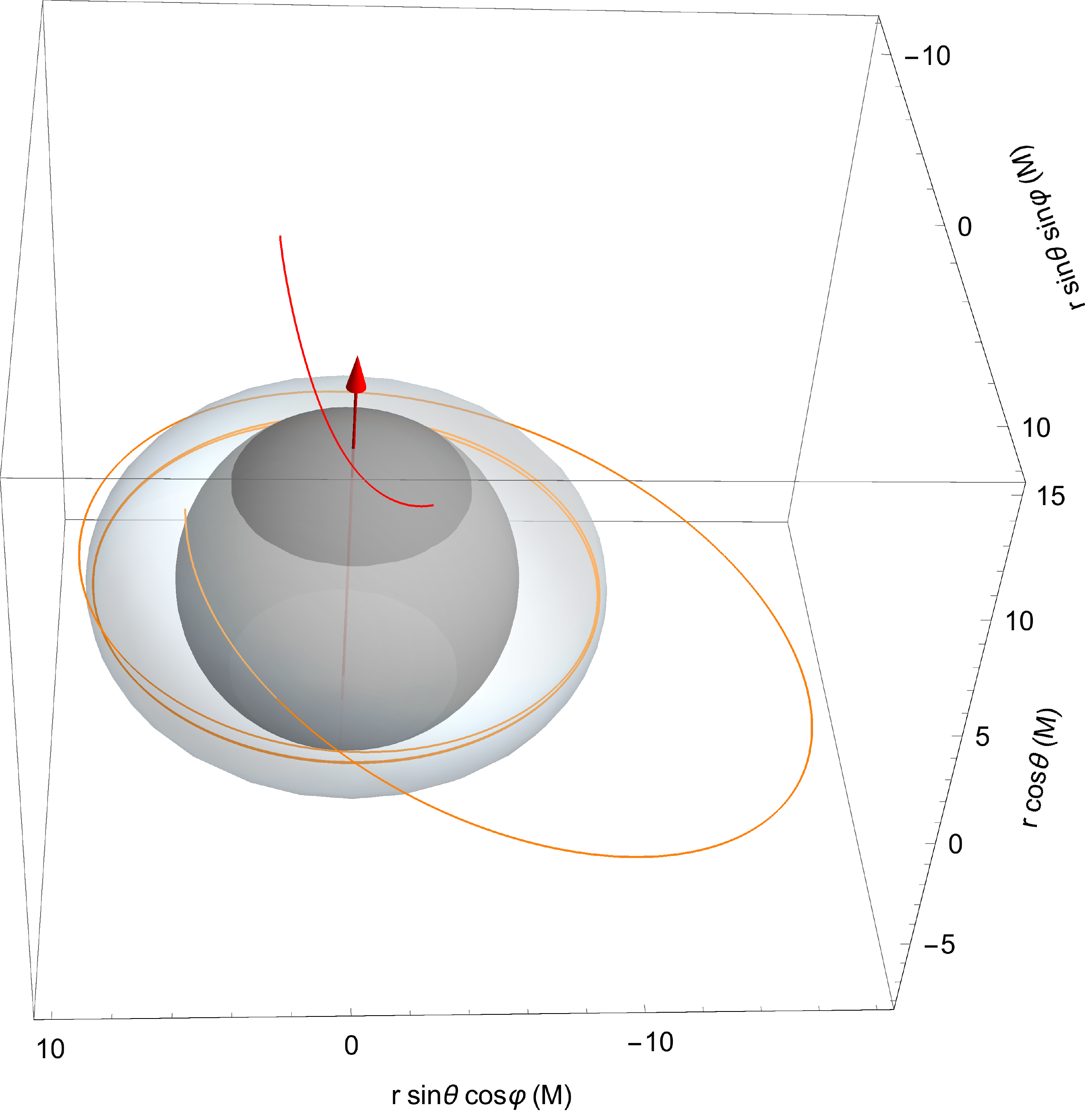}}}
	\caption{Test particles' orbits in the Kerr spacetime plotted in units of $M=1$. For the 2D plot, the continuos black line is the event horizon, the dashed red line is the critical hypersurface, the blue lines are for radial photon with $b=0$, instead the green lines for general photon field with $b\neq0$. 
Left upper panel: the event horizon is at $1.866M$, the spin is $a=-0.5$, the luminosity parameter is $A/M=0.6$, and the critical radius is at $r_{\rm (crit)}=2.71M$. All the test particles have the same initial positions $(r_0,\varphi_0,\alpha_0)=(8M,0,0)$, while the initial velocities are: (1) $\nu_0=0.2$, (2) $\nu_0=0.5$, (3) $(\nu_0,b)=(0.2,1.5)$, and (4) $(\nu_0,b)=(0.2,3.5)$. 
For the 3D plots, the black sphere corresponds to the emitting surface of the NS, and the blue-gray surface denotes the critical hypersurface.
Right upper panel: the event horizon is at $1.866M$, the spin is $a=-0.5$, the luminosity parameter is $A/M=0.8$, and the critical radius is at $r_{\rm (crit)}=5M$. All the test particles have the same initial positions $(r_0,\varphi_0,\alpha_0)=(4M,0,0)$, while the initial velocities are: (1) $\nu_0=0.5$, (2) $\nu_0=0.8$, (3) $\nu_0=0.9$, (4) $(\nu_0,b)=(0.8,0.5)$, and (5) $(\nu_0,b)=(0.8,3.5)$. 	
Left lower panel: Test particle trajectories around a NS of spin $a=0.41$, radius $R_\star=6M$, angular velocity $\Omega_\star=0.031$, and luminosity parameter $A=0.8$, starting at the position $(r_0,\theta_0)=(15M,10^\circ)$ with the initial velocity $\nu_0=0.01$ oriented in the azimuthal corotating direction direction (orange) and oriented radially towards the emitting surface (red).
Right lower panel:Test particle trajectories around a NS of spin $a=0.07$, radius $R_\star=6M$, angular velocity $\Omega_\star=0.005$, and luminosity parameter $A=0.85$, starting at the position $(r_0,\theta_0)=(15M,10^\circ)$ with the initial velocity $\nu_0=0.01$ oriented in the azimuthal corotating direction direction (orange) and oriented radially towards the emitting surface (red).}
	\label{fig:Fig2}
\end{figure*}

In Fig. \ref{fig:Fig2} we display some selected test particle trajectories in 2D and 3D cases in order to show how the PR effect alters the matter motion around a compact object \cite{Bini2009,Bini2011,Defalco20183D,Bakala2019}. It is important to note that in both cases, such effect is strongly sensitive from the initial conditions, therefore it requires that the integration error should be very low, otherwise its propagation could lead to not realistic trajectories. In both models, the test particle has two possible endings, strongly depending on the initial conditions and the parameters defining the geometrical structure and the radiation field: orbiting on the critical hypersurface, or departing at infinity. 

\section{Lagrangian approach: analytical form of the Rayleigh potential}
\label{sec:RF}

\subsection{Mathematical problem}
We consider the following mathematical problem: given the equations of motion of a dissipative system in GR, as Eqs. (\ref{EoM1})--(\ref{EoM6}), of the form $m\boldsymbol{a}(U)=F_{\rm (rad)}(U)^\alpha$, we would like to derive such equations from a principle of least action through the Euler-Lagrange equations
\begin{equation}
\frac{d}{d\tau}\left(\frac{\partial \mathcal{L}}{\partial U_\alpha}\right)-\frac{\partial \mathcal{L}}{\partial X_\alpha}=-\frac{\partial V}{\partial U_\alpha},
\end{equation}
where $\boldsymbol{X}=(t,r,\theta,\varphi)$ and $\boldsymbol{U}=(U^t,U^r,U^\theta,U^\varphi)$, and the unknown functions to be determined are: the Lagrangian function $\mathcal{L}(\boldsymbol{X},\boldsymbol{U})$ (including the kinetic energy and all the conservative and generalised forces) and the Rayleigh dissipative potential $V(\boldsymbol{X},\boldsymbol{U})$ (encompassing the dissipative forces). This is a well posed mathematical problem and it is known in the literature as the inverse problem in the calculus of variations \cite{Santilli1978,Morandi1990,Do2016}. 

Our aim is to derive the analytical form of both functions, without recurring to numerical simulations or codes. We note that the Lagrangian function is already known in the classical literature of GR \cite{Misner1973}, connected with the pure gravitational structure of the background spacetime (i.e., $\boldsymbol{a}=\boldsymbol{0}$), and given by
\begin{equation} \label{eq:lagrangian}
\mathcal{L}(\boldsymbol{X},\boldsymbol{U})=\frac{1}{2}g_{\alpha\beta}U^\alpha U^\beta.
\end{equation}
The challenging issue is of course to determine the analytical form of the Rayleigh potential, connected to the dissipative effects in GR, where generally the dissipative force is highly-nonlinear, because it strongly couples with the curved geometrical background.

\subsection{The method}
The general relativistic PR effect represents the first system in GR, where we have been able to analytically determine the Rayleigh potential \cite{DeFalco2019,DBletter2019,DeFalco2019VE}. To obtain such result, we have exploited two important ideas: the integrating factor to make a differential (semi-basic) one-form closed in its simply connected domain (i.e., exact), and an integration strategy based on the energy dissipated by the system. In this section, we explain into details the followed procedure.

Before to start, we set up some preliminary considerations. The motion of the test particle occurs in $\mathcal{M}$, a simply connected domain (the region outside of the compact object including the event horizon). We denote with $T\mathcal{M}$ the tangent bundle of $\mathcal{M}$, whereas $T^*\mathcal{M}$ stands for the cotangent bundle over $\mathcal{M}$. Let $\boldsymbol{\omega}:T\mathcal{M}\rightarrow T^*\mathcal{M}$ be a smooth differential semi-basic one-form \cite{Libermann1987,MARTINEZ19931,Mestdag2011}, then the radiation force components (\ref{radforce}) can be seen as the components of $\boldsymbol{\omega}$, namely
\begin{equation}
\boldsymbol{\omega}(\boldsymbol{X},\boldsymbol{U})=F_{\rm (rad)}(\boldsymbol{X},\boldsymbol{U})^\alpha \boldsymbol{{\rm d}}X_\alpha. 
\end{equation}

The vertical exterior derivative $\boldsymbol{{\rm d^V}}$ is an operator, whose local expression on a smooth differential semi-basic one-form $\boldsymbol{\beta}=\beta^\alpha \boldsymbol{{\rm d}}X_\alpha$ is given by \cite{Abraham1978,MARTINEZ19931,Mestdag2011}
\begin{equation} \label{eq:vertical_derivative1}
\boldsymbol{{\rm d^V}}\boldsymbol{\beta}= \left(\frac{\partial \beta^\mu}{\partial U_\alpha}-\frac{\partial \beta^\alpha}{\partial U_\mu}\right) \boldsymbol{{\rm d}}X_\alpha\wedge \boldsymbol{{\rm d}}X_\mu,
\end{equation}
whose components are the cross derivatives of $\beta^\alpha$ with respect to $U^\beta$ and $\wedge$ is the wedge product to lift the forms. The differential semi-basic one-form $\boldsymbol{\omega}$ is closed under the vertical exterior derivative $\boldsymbol{{\rm d^V}}$ if $\boldsymbol{{\rm d^V}}\boldsymbol{\omega} = 0$. However, it can be checked that the cross derivatives of (\ref{eq:vertical_derivative1}) are not equal to zero if applied to Eq. (\ref{radforce}). 

Due to the non-linear dependence of the radiation force on the test particle velocity field, the semi-basic one-form turns out to be not exact \cite{Defalco2019}. However, the PR phenomenon exhibits the peculiar propriety according to which $\boldsymbol{\omega}(\boldsymbol{X},\boldsymbol{U})$ becomes exact through the introduction of the integrating factor $\mu =\left(E/\mathbb{E}\right)^2$ \cite{Defalco2019,DBletter2019,DeFalco2019VE}, where $\mathbb{E}=-k_\beta U^\beta$ coincides exactly with $E(U)$\footnote{We preferred to use a different notation for the energy $\mathbb{E}$, instead to continue to call it $E(U)$, to not confuse it with the photon energy $E$, since this variable is very important for what follows.}. 

For the Poincar\'e lemma (generalised to the vertical differentiation) the closure condition and the simply connected domain $\mathcal{TM}$ guarantee that $\mu\boldsymbol{\omega}$ is exact \cite{Martinez1992}. Therefore, it exists a smooth 0-form $V(\boldsymbol{X},\boldsymbol{U})$ such that $-\boldsymbol{{\rm d^V}} V=\mu\boldsymbol{\omega}$, which in local coordinates reads as 
\begin{equation} \label{eq:prim}
\mathbb{F}_{\rm (rad)}(\boldsymbol{X},\boldsymbol{U})^\alpha=-\frac{\partial V(\boldsymbol{X},\boldsymbol{U})}{\partial U_\alpha}.
\end{equation}

Substituting all the occurrences of $\mathbb{E}$ in $F_{\rm (rad)}(\boldsymbol{X},\boldsymbol{U})^\alpha$, see Eq. (\ref{radforce}), we obtain \cite{DBletter2019,DeFalco2019VE} 
\begin{equation} \label{eq:part1}
\mathbb{F}_{\rm (rad)}(\boldsymbol{X},\boldsymbol{U})^\alpha=-k^\alpha \mathbb{E}(\boldsymbol{X},\boldsymbol{U})+\mathbb{E}(\boldsymbol{X},\boldsymbol{U})^2 U^\alpha.
\end{equation}
We consider the velocity derivative operator with respect to the energy $\mathbb{E}$ through the chain rule, having
\begin{equation} \label{eq:trader} 
\frac{\partial\ (\ \cdot\ )}{\partial U_\alpha}=-k^\alpha\ \frac{\partial\ (\ \cdot \ )}{\partial \mathbb{E}}.
\end{equation}
Equation (\ref{eq:prim}) reads explicitly in such case as \cite{DBletter2019,DeFalco2019VE}
\begin{equation} \label{eq:primitive_original}
\mu F_{\rm (rad)}^\alpha=k^\alpha\frac{\partial V}{\partial \mathbb{E}}.
\end{equation}
Considering the scalar product of both members of Eq. (\ref{eq:primitive_original}) by $U_\alpha$, we obtain a differential equation for $V$, which yields at the following integral \cite{DBletter2019,DeFalco2019VE}
\begin{equation} \label{eq:pot_E}
V=-\int \left(\frac{\mu F^\alpha}{\mathbb{E}} \right){\rm d}\mathbb{E}+f(\boldsymbol{X},\boldsymbol{U}),
\end{equation}
where $f(\boldsymbol{X},\boldsymbol{U})$ is constant with respect to $\mathbb{E}$, i.e., $\partial f(\boldsymbol{X},\boldsymbol{U})/\partial \mathbb{E}=0$. Such term can be calculated by using the iterative process of integration of exact differential one-forms, after having rewritten $\mathbb{E}$ as $-k_\alpha U^\alpha$ in order to have all coherently expressed in terms of $(\boldsymbol{X},\boldsymbol{U})$.

Such step is fundamental to understand our strategy, because it permits not only to reduce the calculations from four in terms of the velocity field $\boldsymbol{U}$ to just one in terms of the energy $\mathbb{E}$, but permits to analytically obtain the Rayleigh potential in terms of the energy $\mathbb{E}$ whenever the calculations to obtain $f(\boldsymbol{X},\boldsymbol{U})$ become complicate.

Integrating Eq. (\ref{eq:pot_E}), the final result is \cite{DBletter2019,DeFalco2019VE}
\begin{equation} \label{eq: Rayleigh_potential_final}
V=\tilde{\sigma}\mathcal{I}^2\left[\ln\left(\frac{\mathbb{E}}{E}\right)+\frac{1}{2}\left(U_\alpha U^\alpha+1\right)\right],
\end{equation}
where the constant term $1/2-\ln(E)$ has been determined considering the classical limit, where it is possible to match the general relativistic solution with the classical description \cite{Poynting1903,Robertson1937,DBletter2019,DeFalco2019VE}.  

\subsection{Discussion of the results}
\label{sec:discres}
Our new developed strategy is based on two stages.
\begin{itemize} 
\item[(1)] Use of an integrating factor to make a differential semi-basic one-form exact (closed in its simply connected domain of integration). This approach permits to have more dissipative systems admitting their differential semi-basic one-forms closed, since in GR we deal with dissipative forces highly nonlinear with respect to the velocity field. The closure condition translates in solving a partial differential equation for the integrating factor $\mu$, which in the general relativistic PR case can be easily solved by separation of variables \cite{DeFalco2019VE}; 
\item[(2)] Writing the dissipative force in terms of the dissipated energy $\mathbb{E}$ (see Eq. (\ref{eq:part1})) and passing the derivative operatore from the velocity field to the dissipated energy $\mathbb{E}$ by applying the chain rule (see Eq. (\ref{eq:trader})), we can finally have through some simple algebraic calculations an analytical form of the Rayleigh potential in terms of the dissipated energy $\mathbb{E}$ (see Eq. (\ref{eq:pot_E})). This integral is defined up to a constant function $f(\boldsymbol{X},\boldsymbol{U})$ with respect to the dissipated energy $\mathbb{E}$, i.e., $\partial f(\boldsymbol{X},\boldsymbol{U})/\partial \mathbb{E}=0$. In this term is contained our ignorance on how the dissipative action occurs in the physical system, but at least we know how to act in energetic terms.  
\end{itemize}
\begin{figure*}[t!]
	\centering
	\includegraphics[scale=0.6]{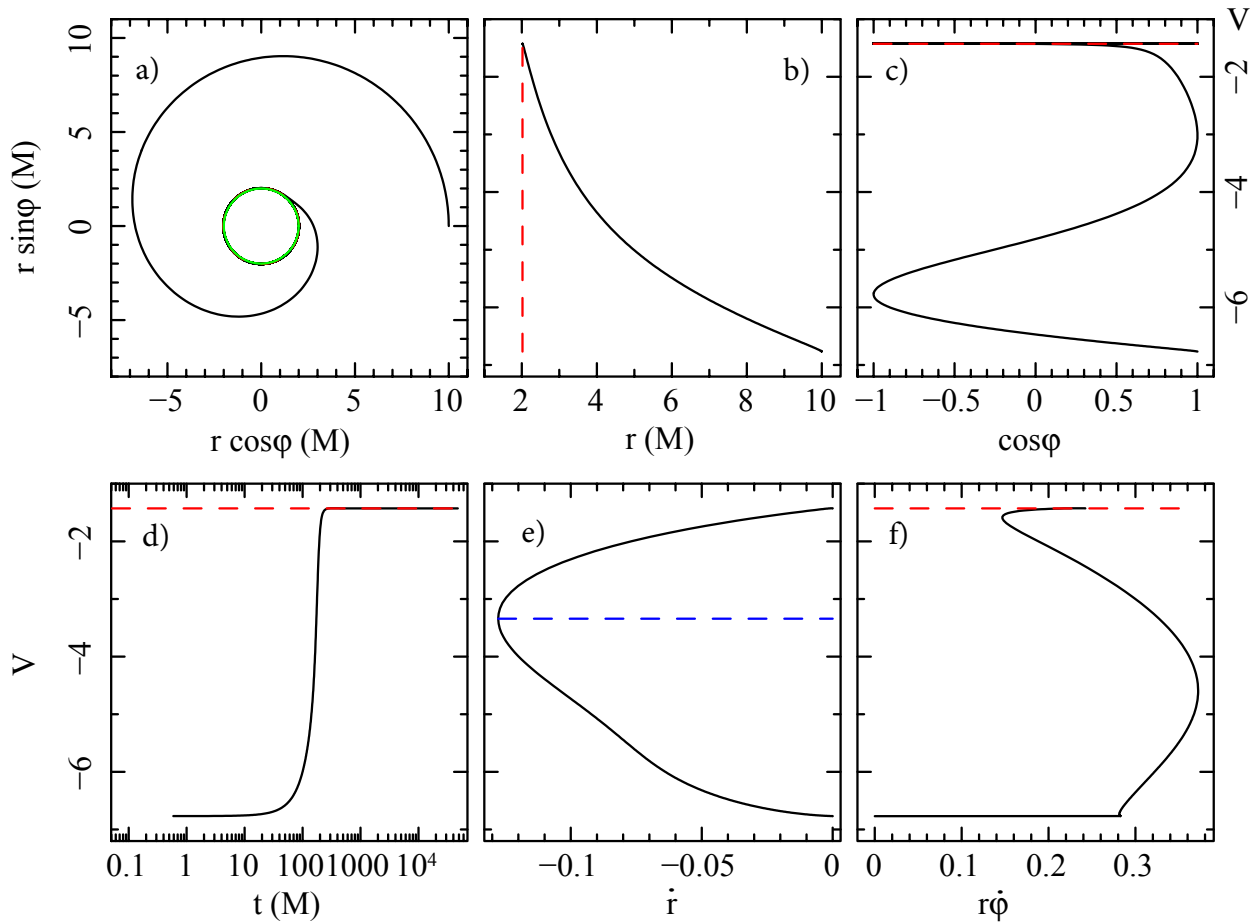}
	\caption{Test particle trajectory with the Rayleigh potential $V$ for mass $M=1$ and spin $a=0.1$, luminosity parameter $A=0.1$ and photon impact parameter $b=1$. The test particle moves in the spatial equatorial plane with initial position $(r_0,\varphi_0)=(10M,0)$ and velocity $(\nu_0,\alpha_0)=(\sqrt{1/10M},0)$. a) Test particle trajectory spiralling towards the BH and stopping on the critical radius (red dashed line) $r_{\rm (crit)}=2.02M$. The continuous green line is the event horizon radius $r^+_{\rm (EH)}=1.99M$. Rayleigh potential versus b) radial coordinate, c) azimuthal coordinate, d) time coordinate, e) radial velocity, and f) azimuthal velocity. The blue dashed line in panel e) marks the minimum value attained by the radial velocity, corresponding to $\dot{r}=-0.13$.}
	\label{fig:Fig4}
\end{figure*}

It is important to note that $\mathbb{E}$ is considered as the dissipated energy, because we can see our radiation field made by photon-bullets, which are shot against the test particle-target. Faster the test particle moves, less are the photons absorbed, or in other words, more are the photons dissipated. Indeed, if the test particle is at rest, $\mathbb{E}=E$, which is the maximum attained energy; while instead if the test particle moves close to a speed of light, $\mathbb{E}=0$ is the minimum reachable energy \cite{DeFalco2019VE}. 

The analytical form of the Rayleigh potential related to the general relativistic PR effect is very important because it represents the first example in the GR literature. The Rayleigh potential (\ref{eq:pot_E}) is a valuable tool to investigate the proprieties of the general relativistic PR effect and more in general the radiation processes in high-energy astrophysics. In addition, such function entails two main important consequences:
\begin{itemize}
\item the Rayleigh potential contains a logarithm of the energy, which, in view of the interpretation of the dissipated energy $\mathbb{E}$, can be physically interpreted as the absorbed energy from the test particle, while the other function $U^\alpha U_\alpha$ represents the re-emission process, which is in agreement with the underlying hypothesis of the PR effect model, i.e., the test particle behaves as an ideal black body in thermal equilibrium, which re-emits radiation at constant rate, isotropically, and independent from the velocity field $U^\alpha$ \cite{DeFalco2019VE}.
\item the Rayleigh potential permits to directly connect  the theory with the observations. This last sentence can be better understood by looking at Fig. \ref{fig:Fig4}, where in panel $a)$ the test particle trajectory is displayed (i.e., what can be observed) and in panels $b)-f)$ the Rayleigh potential in terms of the coordinates $r,\varphi,t,\dot{r},r\dot{\varphi}$, respectively (i.e., what derives from the theory). In other words, detecting the test particle motion, it is possible to infer the analytic structure of the Rayleigh potential; viceversa by using different functional form of the Rayleigh potential to investigate other kinds of radiation processes in high-energy astrophysics, it can be possibile to numerically simulate the test particle trajectory (see Ref. \cite{DeFalco2019VE}, for details).  
\end{itemize}

\section{Dynamical system approach: stability of the critical hypersurfaces}
\label{sec:stbch}
The general relativistic PR effect can be also analysed under the dynamical system point of view for proving the stability of the critical hypersurfaces. To this end, we consider only those initial conditions, where the test particle ends its motion on the critical hypersurfaces without escaping at infinity. It is not possible to formally characterise these configurations, because the general relativistic PR effect models show a sensitive dependence from the initial data\footnote{A dynamical system is defined to be sensitive dependent from the initial data, when tiny perturbations to the initial data give extremely different future behaviors.}. Once the stability issue has been proved, it immediately follows that the critical equatorial ring is a stable attractor (region where the test particle is attracted for ending its motion), and the whole critical hypersurface is a basin of attraction \cite{Defalco2019ST}.  

\subsection{Previous approach: linearization theory}
\label{sec:old}
In the previous approach pursued by Bini and collaborators, they have been able to prove such statement only in the Schwarzschild case within the linear stability theory (see Appendix in Ref. \cite{Bini2011}). This method relies on the linearization of a dynamical system $\boldsymbol{\dot{x}}=\boldsymbol{f}(\boldsymbol{x})$ towards the critical point $\boldsymbol{x_0}$ obtaining thus $\boldsymbol{\dot{x}}\approx\boldsymbol{A}\cdot\boldsymbol{x}+O(|\boldsymbol{x}|^2)$, where $\boldsymbol{A}=[\nabla_{\boldsymbol{x}}\boldsymbol{f}(\boldsymbol{x})]_{\boldsymbol{x_0}}$. Then searching for the eigenvalues of the matrix $\boldsymbol{A}$, and checking whether they are all negatives, we can finally obtain that the dynamical system is stable toward the critical point $\boldsymbol{x_0}$. At glance, this approach seems theoretically simple, but practically it implies computationally-expensive calculations (especially in the Kerr case).

\subsection{New approach: Lyapunov theory}
\label{sec:new}
We have introduced a new, simpler in term s of calculations, and more physical approach based on the Lyapunov theory \cite{Robert2004}. The dynamical system (\ref{EoM1})--(\ref{EoM6}), $\dot{\boldsymbol{x}}=\boldsymbol{f}(\boldsymbol{x})$, is defined in the domain $\mathcal{D}$ (spacetime outside the compact object including the event horizon), and we call with $\mathcal{H}$ the critical hypersurface. Let $\Lambda=\Lambda(\boldsymbol{x})$ be a smooth and real valued function, continuously differentiable in all points of $\mathcal{D}$, then $\Lambda$ is a Lyapunov function for $\dot{\boldsymbol{x}}=\boldsymbol{f}(\boldsymbol{x})$ if it satisfies the following conditions:
\begin{eqnarray}  
{\rm (I)}&&\quad \Lambda(\boldsymbol{x})>0,\quad \forall \boldsymbol{x}\in \mathcal{D}\setminus\mathcal{H};\label{eq:lia1}\\
{\rm (II)}&&\quad \Lambda(\boldsymbol{x_0})=0,\quad \forall \boldsymbol{x_0}\in \mathcal{H};\label{eq:lia2}\\ 
{\rm (III)}&&\quad \dot{\Lambda}(\boldsymbol{x})\equiv\nabla\Lambda(\boldsymbol{x})\cdot \boldsymbol{f}(\boldsymbol{x})\le0 ,\quad \forall \boldsymbol{x}\in \mathcal{D}\label{eq:lia3}.
\end{eqnarray}
Once a Lyapunov function $\Lambda$ is determined for all points belonging to the critical hypersurface $\mathcal{H}$, a theorem due to Lyapunov assures that $\mathcal{H}$ is stable \cite{Robert2004}. The conditions $(I)$ and $(II)$ are very easy to be proved, while the last condition $(III)$ requires to perform some calculations, that anyway can be easily carried out by hand.

Such approaches has the great advantage to study the behavior of a dynamical system without knowing its analytical solution or performing any approximation. Nevertheless, some limits should be also taken into account, like: there is any fixed rule or recipe to determine it, but sometimes only physical intuitions can help; it is not unique, but it could be possible to determine more than one or sometimes can even not exist. 

\subsubsection{Three Lyapunov functions for the general relativistic PR effect}
\begin{figure*}[th!]
	\centering
	\vbox{
	\hbox{\hspace{0cm}
		\includegraphics[scale=0.32]{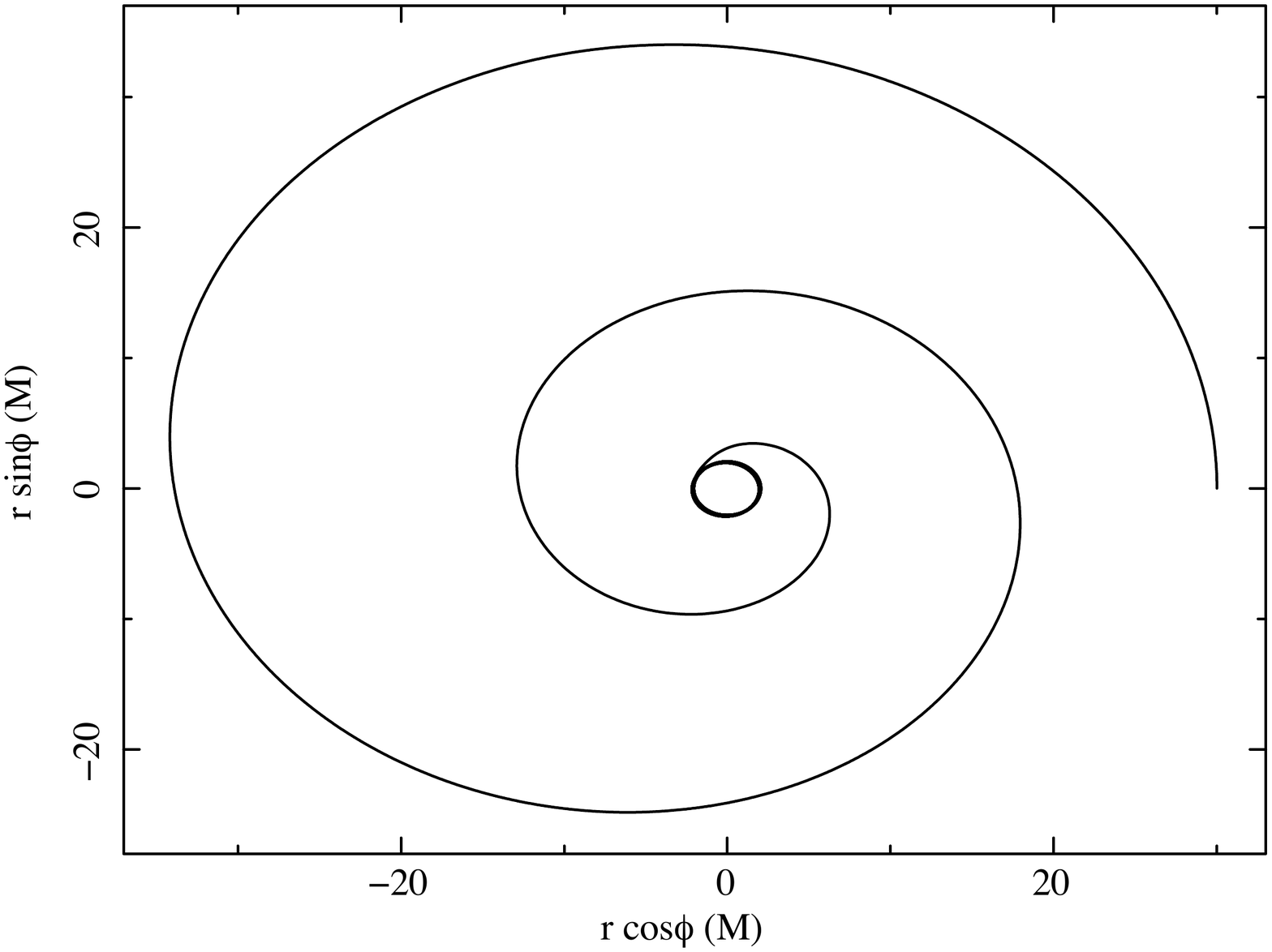}
		\hspace{-1 cm}
		\includegraphics[scale=0.32]{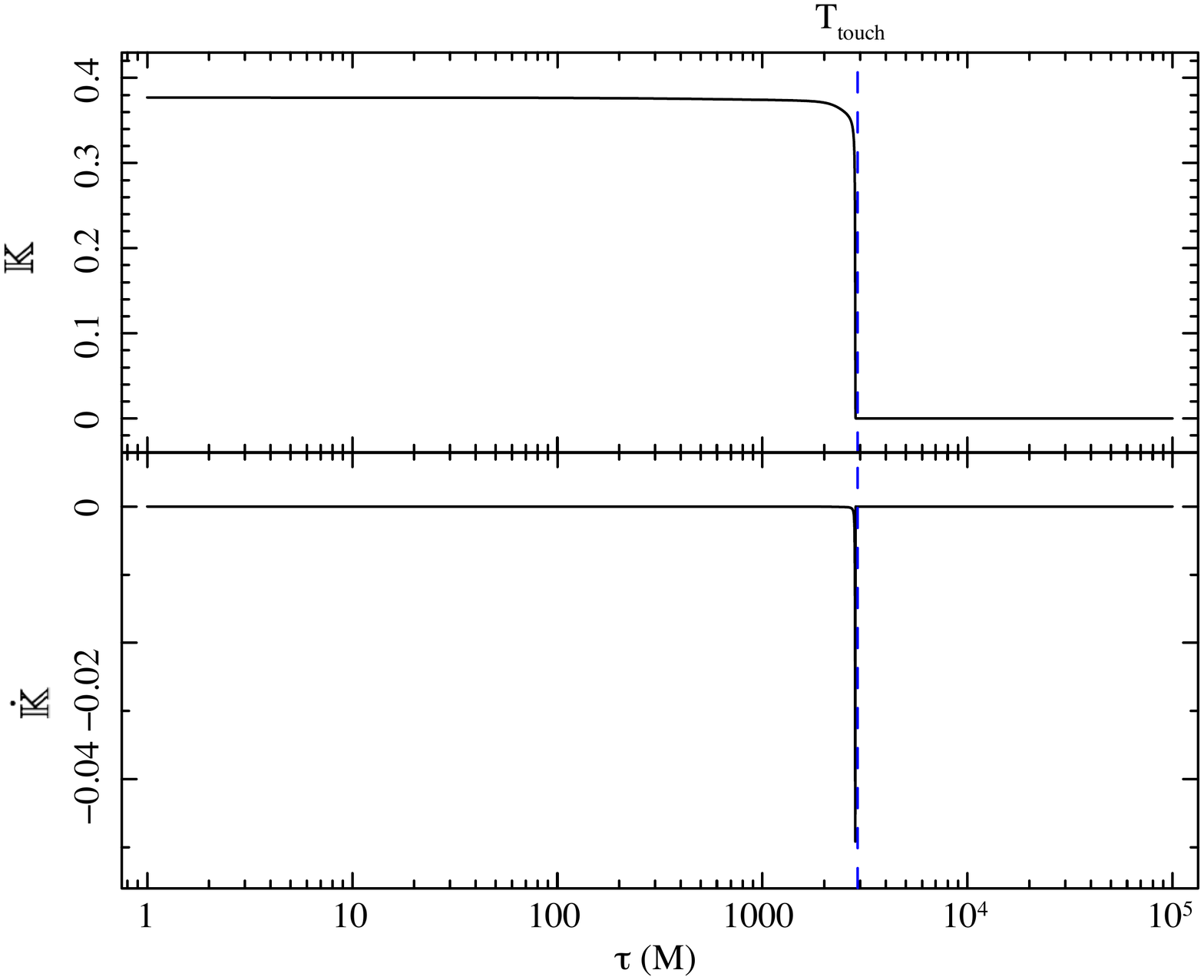}}
		\vspace{-0.3 cm}
	\hbox{\hspace{0cm}
		\includegraphics[scale=0.32]{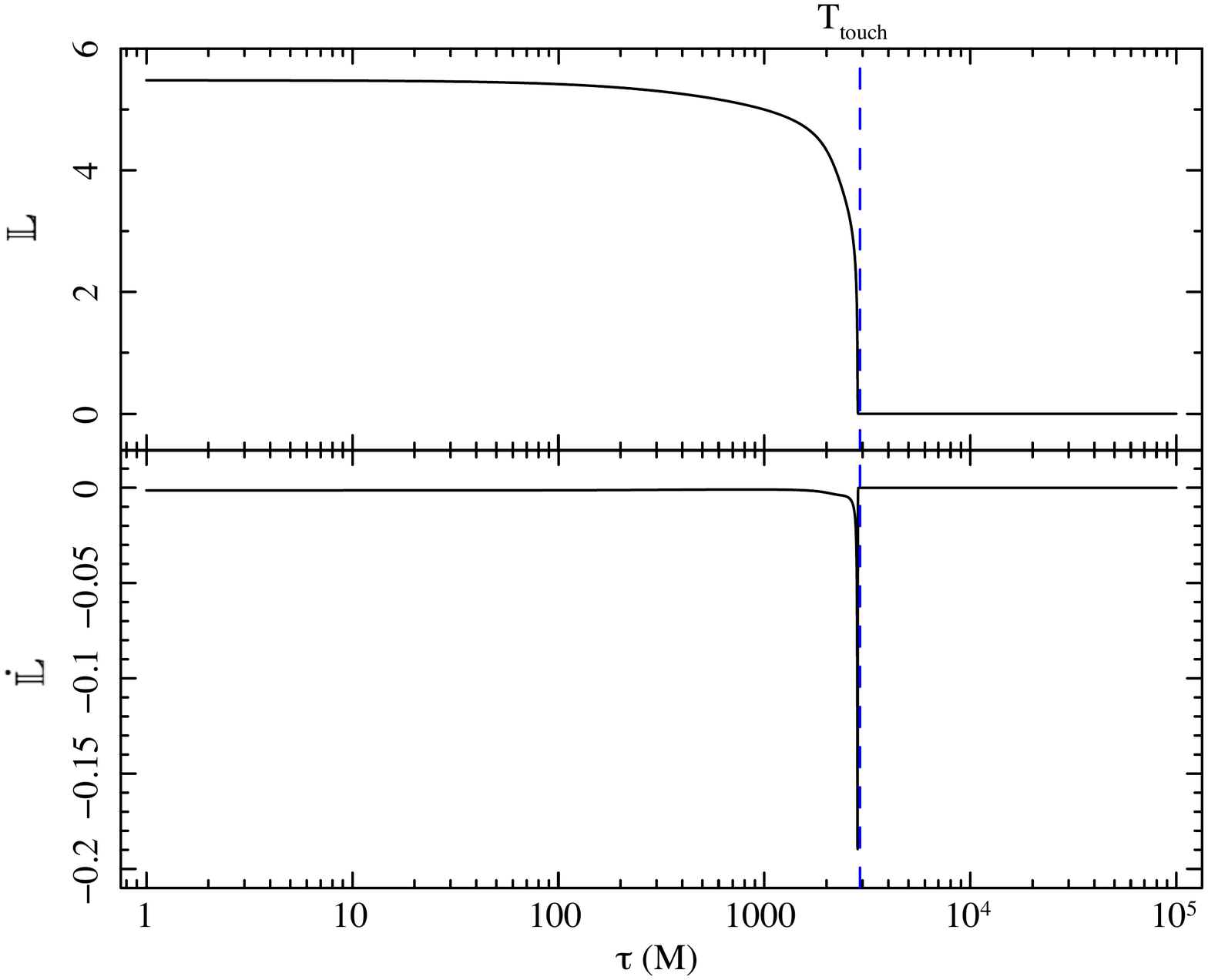}
		\hspace{-1 cm}
		\includegraphics[scale=0.32]{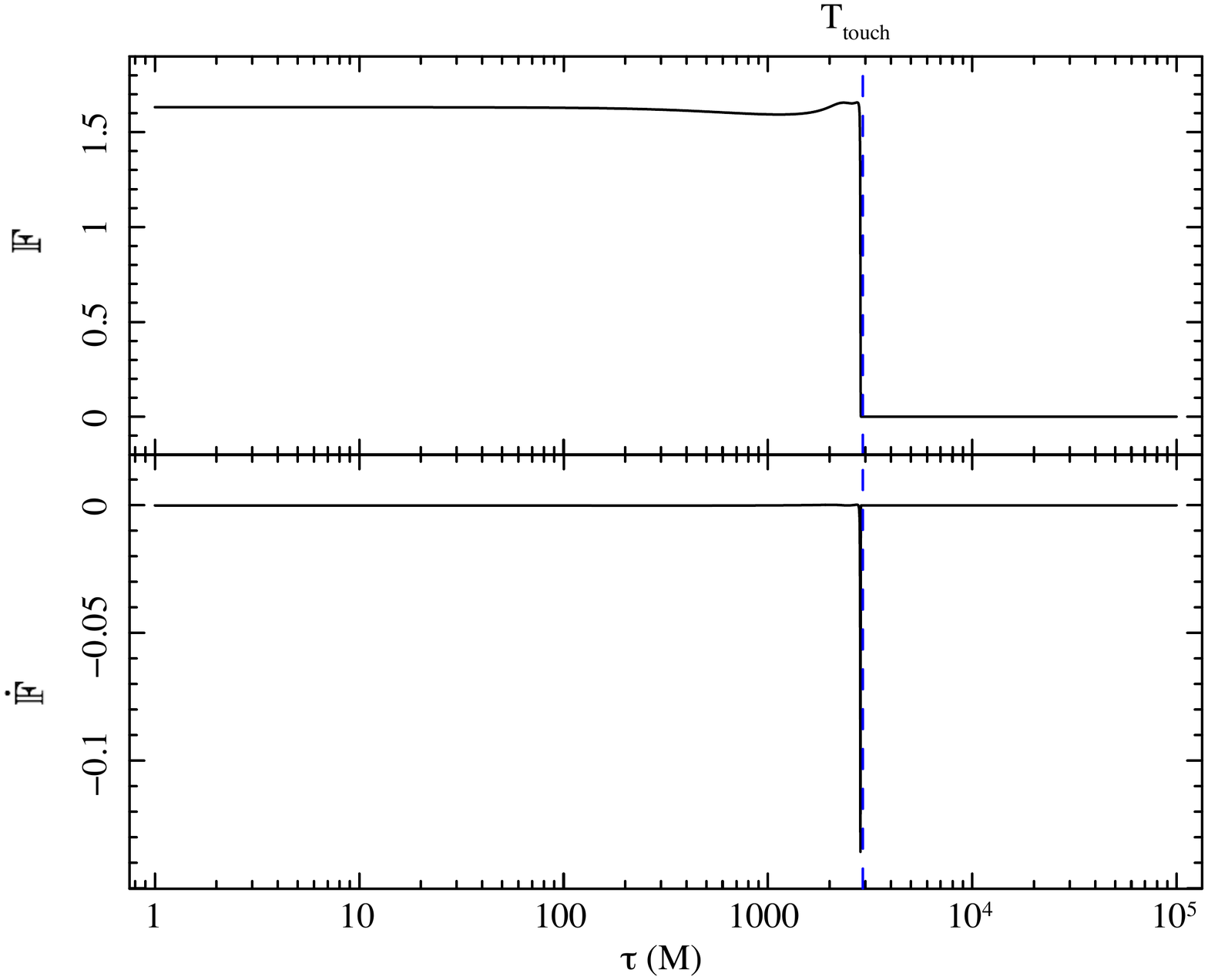}}}
	\caption{We show a test particle orbit and the related three Lyapunov functions. Upper left panel: test particle moving around a rotating compact object with mass $M=1$, spin $a=0.3$, luminosity parameter $A=0.2$, and photon impact parameter $b=0$. The test particle starts its motion at the position $(r_0,\varphi_0)=(30M,0)$ with velocity $(\nu_0,\alpha_0)=(\sqrt{M/r_0},0)$. The critical hypersurface is a circle with radius $r_{\rm (crit)}=2.07M$. The energy (see Eqs. (\ref{eq:LF1}) and (\ref{eq:DLF1}), and upper right panel), the angular momentum (see Eqs. (\ref{eq:LF2}) and (\ref{eq:DLF2}), and lower left panel), and the Rayleigh potential (see Eqs. (\ref{eq:LF3}) and (\ref{eq:DLF3}), and lower right panel) together with their $\tau$-derivatives are all expressed in terms of the proper time $\tau$. The dashed blue lines in all plots represent the proper time $T_{\rm touch}$ at which the test particle reaches the critical hypersurface and it amounts to $T_{\rm touch}=2915M$.}
	\label{fig:Fig3}
\end{figure*}

For the general relativistic PR effect, three different Lyapunov functions have been determined. An idea of the proof that they are Lyapunov functions is based on expanding all the kinematic terms of the equations of motion with respect to the radius, estimating thus their magnitude, and then considering that the test particle's orbit is confined in a bounded box (since by hypothesis we consider only those configurations reaching the critical hypersurface), see Ref. \cite{Defalco2019ST}, for further details.
\begin{itemize}
\item \emph{The relative classical mechanical energy} of the test particle with respect to the critical hypersurface measured in the ZAMO frame is 
\begin{equation} \label{eq:LF1}
\mathbb{K}=\frac{m}{2}\left|\nu^2-\nu^2_{\rm crit}\right|+(A-M)\left(\frac{1}{r}-\frac{1}{r_{\rm crit}}\right),
\end{equation}
where $\nu_{\rm crit}(\theta)=[\cos\beta]_{r=r_{\rm crit}(\theta)}$, which includes as a particular case the velocity $\nu_{\rm eq}=[\cos\beta]_{r=r_{\rm crit}(\pi/2)}$ in the equatorial ring. Its derivative is 
\begin{equation}  \label{eq:DLF1}
\begin{aligned}
\dot{\mathbb{K}}&=m\ {\rm sgn}\left(\nu^2-\cos^2\beta\right)\left[\nu\frac{d\nu}{d\tau}-\cos\beta\frac{d (\cos\beta)}{d \tau}\right]\\
&-\frac{A-M}{r^2}\dot{r}.
\end{aligned}
\end{equation}
where ${\rm sgn}(x)$ is the signum function.
\item \emph{The relative classical angular momentum} of the test particle measured in the ZAMO frame is 
\begin{equation}  \label{eq:LF2}
\begin{aligned}
\mathbb{L}=m(r\nu\sin\psi\cos\alpha-r_{\rm crit}\nu_{\rm crit}).
\end{aligned}
\end{equation}
Its derivative is given by
\begin{equation}  \label{eq:DLF2}
\begin{aligned}
\dot{\mathbb{L}}&=m\ \left[-\dot{r}_{\rm crit}\nu_{\rm crit}-r_{\rm crit}\frac{d(\nu_{\rm crit})}{d\tau}\right.\\
&\left.+r\frac{d\nu}{d\tau}\cos\alpha\sin\psi+\nu(\dot{r}\cos\alpha\sin\psi\right.\\
&\left.-r\sin\alpha\sin\psi\ \dot{\alpha}+r\sin\alpha\cos\psi\ \dot{\psi})\right].
\end{aligned}
\end{equation}
\item \emph{The relative general relativistic Rayleigh dissipation function} is (see Sec. \ref{sec:RF} for details)
\begin{equation}
\mathbb{F}=\tilde{\sigma}\mathcal{I}^2\left[\lg\left(\frac{\mathbb{E_{\rm crit}}}{E_p}\right)-\lg\left(\frac{\mathbb{E}}{E_p}\right)\right],
\end{equation}
where 
\begin{equation}  \label{eq:LF3}
\begin{aligned}
\mathbb{E}&\equiv-k_\alpha U^\alpha\\
&=\gamma \frac{E_p}{N}(1+bN^\varphi)[1-\nu\sin\psi\cos(\alpha-\beta)].
\end{aligned}
\end{equation}
$\mathbb{E_{\rm crit}}$ is the energy $\mathbb{E}$ evaluated on the critical hypersurface, given by 
\begin{equation} 
\begin{aligned}
\mathbb{E_{\rm crit}}&=[\mathbb{E}]_{r=R_\star,\alpha=0,\pi,\psi=\pm\pi/2,\nu=\nu_{\rm crit}}\\
&=\frac{E_p|(\sin\beta)_{\rm crit}|}{N_{\rm crit}}(1+bN^\varphi_{\rm crit}).
\end{aligned}
\end{equation}
Its derivative is
\begin{equation}  \label{eq:DLF3}
\begin{aligned}
\dot{\mathbb{F}}&=\tilde{\sigma}\dot{(\mathcal{I}^2)}\left[\lg\left(\frac{\mathbb{E_{\rm crit}}}{E_p}\right)-\lg\left(\frac{\mathbb{E}}{E_p}\right)\right]\\
&+\tilde{\sigma}\mathcal{I}^2\left[\frac{\dot{\mathbb{E}}_{\rm crit}}{\mathbb{E_{\rm crit}}}-\frac{\dot{\mathbb{E}}}{\mathbb{E}}\right].
\end{aligned}
\end{equation}
\end{itemize}
In Fig. \ref{fig:Fig3} we show the trajectory of a test particle orbiting in the equatorial plane around a BH and influenced by a radiation field together with the general relativistic PR effect. The test particle ends its motion on the critical hypersurface, which is the configuration we focussed on. In the other panels of Fig. \ref{fig:Fig3}, the trend of the three proposed functions have been displayed (i.e., $\mathbb{K},\ \mathbb{L},\ \mathbb{F}$), together with their derivatives (i.e., $\dot{\mathbb{K}},\ \dot{\mathbb{L}},\ \dot{\mathbb{F}}$), to graphically prove that they verify the three proprieties to be Lyapunov functions. 

We note that the the first two Lyapunov functions (mechanical energy and angular momentum) are classical definition, while the third example (Rayleigh dissipation function) is a pure general relativistic case. The former two functions are not in contradiction with the latter, rather they are very useful to substantially reduce the calculations with respect to their general relativistic versions. In addition, it must be said that another great advantage of Lyapunov theory is that the Lyapunov functions should not need to have a physical meaning, they can be also mathematical function, which should respect the three conditions stated above, so that they can prove the stability of the critical hypersurfaces.

\section{Conclusions}
\label{sec:end}
In this work we have presented three different and complementary approaches to study the general relativistic PR effect. They can be summarised as: 
\begin{itemize}
\item working on the model by improving it in its ingenuous aspects (see Sec. \ref{sec:3D}). Indeed, we have improved the PR effect model from the 2D equatorial plane to the 3D space in Kerr geometry. The emitted photons are in general parametrized by two impact parameters $(b,q)$, since we are in the 3D case. However, imposing that the radiation field must lie in the equatorial plane of the ZAMO frame (even at infinity), we not only reduce the radiation field to only one parameter, but we also develop an astrophysical coherent model. The resulting equations of motion represent a system of six coupled ordinary and highly nonlinear differential equations of first order, see Eqs. (\ref{EoM1})--(\ref{EoM6}). Such dynamical system admits the existence of a critical hypersurface, regions where the gravitational attraction is balanced by the radiation force. The test particle can end its motion on it, moving over there stably (see Sec. \ref{sec:critc_rad}). The main difference of the 3D model with the 2D case is in having the phenomena of latitudinal drift and suspended orbits.
\item the general relativistic PR effect can be seen as a dissipative system (see Sec. \ref{sec:RF}). The equations of motion can be treated within the theory of inverse problem of calculus of variations, where the unknown functions are the Lagrangian and the Rayleigh potential. The former is easily found (\ref{eq:lagrangian}), while for the latter we have developed a strategy to determine it analytically (\ref{eq: Rayleigh_potential_final}). Such approach is based first on making a differential semi-basic one-form exact through the introduction of an integrating factor, and then to substantially reduce the calculations for obtaining its analytic expression at least in terms of the dissipated energy (\ref{eq:pot_E}). The full complete analytical form of the Rayleigh potential permitted to discover a new functional class related to absorption processes in high-energy astrophysics (see Sec. \ref{sec:discres}), and to develop a strategy to closely relate observations and theory (see Fig. \ref{fig:Fig4}).    
\item another perspective to analyse the general relativistic PR effect is under the dynamical system point of view (see Sec. \ref{sec:stbch}). We have proved the stability of the critical hypersurfaces by introducing a new method. The previous approach was based on the linearization theory, where the calculations revealed to be very demanding (see Sec. \ref{sec:old}). Therefore, we have thought to introduce a more straightforward method within the Lyapunov theory (see Sec. \ref{sec:new}). The determination of three different Lyapunov functions (i.e., classical mechanical energy (\ref{eq:LF1}) and angular momentum (\ref{eq:LF2}) of the test particle and general relativistic Rayleigh potential (\ref{eq:LF3})) permitted to prove in an equivalent way and under different physical aspects the stability of the critical hypersurfaces.
\end{itemize}

As future projects, we plan to further investigate the results obtained in the three different approaches, namely: (1) improve the actual theoretical assessments employed to model the radiation field in some elementary aspects, like: the momemntum-transfer cross section will be not anymore constant, but it will depend on the angle and frequency of the incoming radiation field, the radiation field is not emitted anymore by a point-like source, but from a finite extended source; (2) generalise the strategy for finding the analytical form of the Rayleigh potential to other dissipative systems in GR, like for example the gravitational wave theory; (3) apply the Lyapunov functions to all the extensions of the general relativistic PR effect model with the due modifications.

Anyway we would like to find other alternative approaches to the same problem in order to extract several other interesting information, and in the same time to develop new formalisms, which can be applied to other dissipative systems in GR. We would like also to apply the general relativistic PR effect model to describe some astrophysical problems, like: accretion disk model, type-I X-ray burst, photospheric radius expansion. 

\subsection*{Funding and conflict of interest}
The author thanks the Silesian University in Opava and Gruppo Nazionale di Fisica Matematica of Istituto Nazionale di Alta Matematica for support. The results contained in the present paper have been partially presented at the conference \emph{WASCOM 2019}, Maiori (Italy).

\bibliography{references}
\end{document}